\documentclass[a4paper,11pt]{article}
\pdfoutput=1 % if your are submitting a pdflatex (i.e. if you have
\usepackage{jinstpub} % for details on the use of the package, please
\usepackage{lineno}
\usepackage[english]{babel}
\usepackage{siunitx}
\usepackage{float}

\title{Scintillation detectors with silicon photomultiplier readout in a dilution refrigerator at temperatures down to 0.2 K}

\author[1]{J. Zhang,}
\author{D. Goeldi,}
\author{R. Iwai,}
\author{M. Sakurai,}
\author[1]{and A. Soter, \note{Corresponding author.}}
\affiliation{Institute for Particle Physics and Astrophysics, ETH Z\"urich,\\8093 Z\"urich, Switzerland}
\emailAdd{zhangje@phys.ethz.ch}
\emailAdd{asoter@phys.ethz.ch}

\abstract{
We are developing a novel high-brightness atomic beam, comprised of a two-body exotic atom called muonium (M $ = \mu^+ + e^-$), for next-generation atomic physics and gravitational interaction measurements. This M source originates from a thin sheet of superfluid helium (SFHe), hence diagnostics and later measurements require a detection system which is operational in a dilution cryostat at temperatures below \SI{1}{\K}. In this paper, we describe the operation and characterization of silicon photomultipliers (SiPMs) at ultra-low temperatures in SFHe targets. We show the temperature dependence of the signal shape, breakdown voltage, and single photon detection efficiency, concluding that single photon detection with SiPMs below  \SI{0.85}{\K} is feasible. Furthermore, we show the development of segmented scintillation detectors, where 16 channels at \SI{1.7}{\K} and one channel at \SI{170}{\milli \K} were commissioned using a muon beam.
}

\keywords{Cryogenic detectors; Photon detectors for UV, visible, and IR photons (solid-state); Solid state detectors}

\arxivnumber{2203.15631}

\begin{document}
\maketitle
\flushbottom

\section{Introduction}
\label{sec:intro}
The operation of silicon photomultipliers (SiPMs) at very low temperatures has the potential to improve detector systems for various experiments conducted at cryogenic temperatures that require single photon detection or sensitivity in the vacuum ultraviolet (VUV) range. In rare event searches, SiPMs are promising candidates to replace photomultiplier tubes (PMTs) to detect VUV scintillation photons produced by particle interactions with liquid xenon and argon. Examples are the dark matter experiments DARWIN \cite{Aalbers2016} or DarkSide-20k \cite{darkside}, nEXO \cite{Albert2018} looking for neutrino-less double beta decay, or MEGII \cite{Ogawa2017} searching for the charged lepton flavour violating decay of muons. The characterization of SiPMs has thus been done at room temperature \cite{Gallina2019} and down to liquid-nitrogen temperatures (\SI{77}{\K}) \cite{Falcone2015, Iwai2019}. 

The detection of scintillation light at even lower temperatures is of interest for several experiments conducted at the Paul Scherrer Institute (PSI), like the muCool experiment, where positive muons are moderated in helium gas targets at $\sim$ \SI{4}{\K} to \SI{10}{\K} temperatures  \cite{Antognini2021, antognini2020}. In a new experiment, we are developing a high-quality atomic muonium (M $ = \mu^+ + e^-$) source, produced in a \SI{150}{\milli \K} to \SI{200}{\milli \K} superfluid helium (SFHe) sheet for a direct measurement of the gravitational interaction of second-generation, leptonic antimatter \cite{Soter2021, Kirch2014}. These experiments utilize scintillators at cryogenic temperatures. However, the scintillation light is typically transported to a room-temperature environment using plastic fibres before being detected by photodetectors. The application of cryogenic SiPMs in the cryogenic environment could reduce the complexity of such experiments and allow for finer detector geometries \cite{soter2014segmented}. Furthermore, single photon detection capabilities at ultra-low temperatures is of increasing interest in a variety of applications in quantum information science, where SiPMs have been mainly used at room temperature \cite{Eisaman2011}. 

In this work we show that SiPMs can be operated to detect single photons inside a dilution cryostat with limited cooling power at temperatures below \SI{1}{\K}, a behavior which was not expected due to the overwhelming afterpulsing measured below the advertised lowest operational ranges ($\sim$\SI{70}{\K}). The temperature dependence of the signal shape, breakdown voltage, and photon detection efficiency are presented as well as the operating range of the SiPM for single photon detection. Finally, we show how the SiPM can be used in a scintillation detector system with multiple channels, capable of operating in a dilution cryostat with the coldest detector operating at \SI{170}{\milli \K}.

\section{Experimental setup}
\label{sec:experimental setup}

\begin{figure}[hb!]
\centering
\includegraphics[width=.95\textwidth,trim=0 0 0 0,clip]{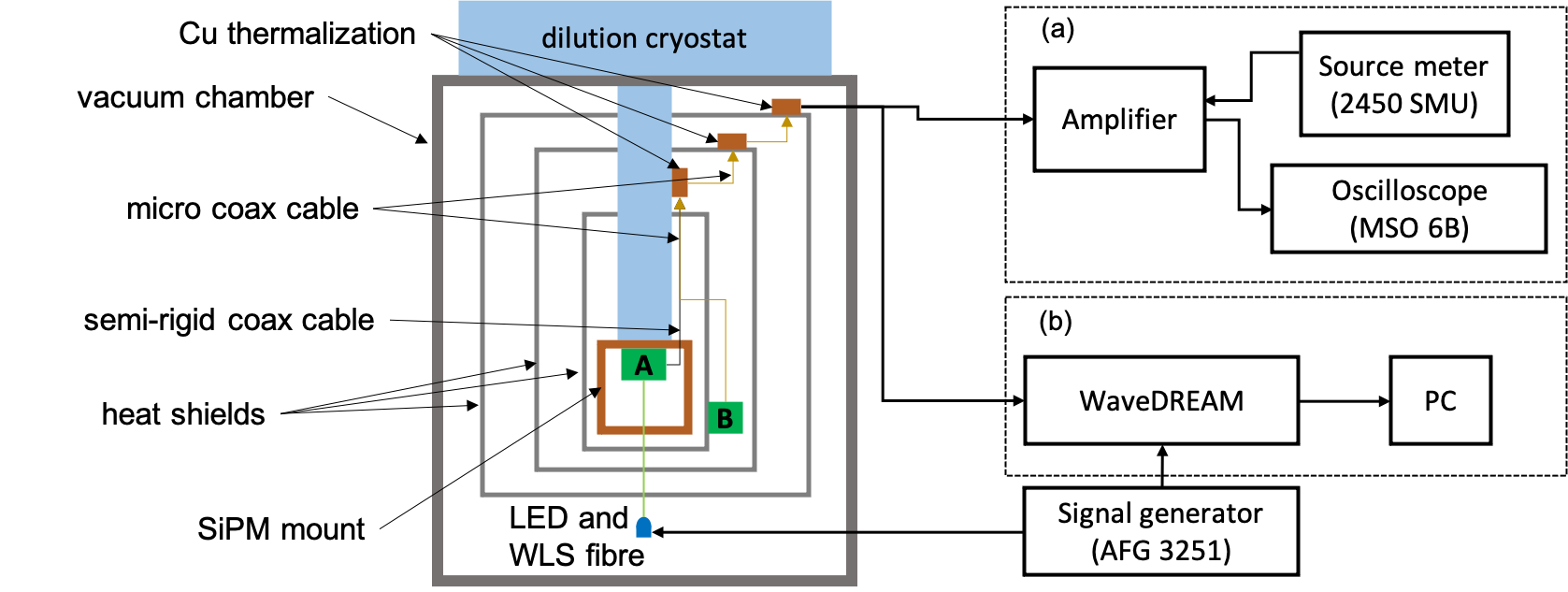}
\caption{\label{fig:Setup}
Schematic view of the experimental setup and the data acquisition system. For the temperature dependent characterization, a SiPM was mounted within a copper structure (position A) in contact with the cold plate. Up to 16 additional cryogenic detectors can be positioned on the first heat shield (position B). The signal from the SiPM was then either amplified and read out using an oscilloscope (a), or directly fed to the WaveDREAM board (b).}
\end{figure}

The device studied in this work was the S13370 (VUV4) SiPM from Hamamatsu, which was originally developed for the detection of VUV photons from liquid xenon and argon targets. Specific features are the absence of the protection layer to improve VUV sensitivity and the utilization of metallic quenching resistors making the resistance less temperature dependent.
To study the performance of SiPMs at ultra-low temperatures a custom-made dilution cryostat was used \cite{VandenBrandt1990}. A sketch of the experimental setup is shown in figure \ref{fig:Setup}. A SiPM was thermalized in an oxygen-free high-thermal-conductivity copper support in contact with the cold plate (position A), which is  protected from thermal radiation by three layers of aluminium heat shields. Two temperature sensors (LakeShore CX-1010) were placed on the copper support close to the SiPM to monitor the temperature. Additional detectors situated at higher temperatures could be placed on acrylic supports on the first heat shield (position B), where the temperature was also monitored using the same type of sensor.

Due to the limited cooling power of the cryostat, the heat load from the signal cables needed to be minimized.  We installed 18 channels of $\sim$\SI{5}{\m} long micro coaxial cable (50 $\Omega$, 38 AWG tinned copper alloy, \SI{0.4}{\milli \m} outer sleeve diameter), which we chose as a compromise between heat load and signal quality. Each channel was thermalized at three temperature stages in the cryostat and terminated at \SI{2.4}{\K} in a row of MMCX connectors. From there, 16 channels were connected with the same wires to the SiPMs at position B, while a semi-rigid coaxial cable (50 $\Omega$ impedance) led to the SiPM at position A.

The SiPM characterization at cryogenic temperatures was done using a sensor with 3 $\times$ 3 \si{\mm^2} active area and \SI{75}{\um} pixel pitch, placed at position~A. The sensor faced the polished end of a wavelength shifting fibre (Kuraray Y-11) with peak emission wavelength at $\lambda$ = 476 nm, which at the other end was coupled to a LED (Thorlabs TO-18, 450 nm, 7nW) placed in the vacuum chamber of the cryostat at room temperature. A signal generator (Tektronix AFG 3251) was used to bias the LED with a square pulse (\SI{100}{\kilo \Hz} rate, \SI{10}{\nano \s} width, \SI{-2.7}{\V} amplitude) and to simultaneously trigger the data acquisition system.

We used two different setups for the measurement. In setup (a), a source measurement unit (Keithley 2450 SMU) was used to provide the bias voltage and to measure the resulting current. The signal from the SiPM was amplified and then read out by a 20 GS/s oscilloscope (Tektronix MSO6B). Amplification is done using a MAR-amplifier, developed at PSI for fast photo-diode applications, with a voltage gain of 40 dB (37 dB) at 100 kHz (750 MHz) and input impedance of 50 $\Omega$.
In setup (b), the raw signal of the SiPM was read out using the WaveDREAM board, based on DRS4 digitizers developed at PSI, which provided the bias voltage, amplified the input signal, and digitized the waveforms \cite{Galli:2019nmv}.

\section{SiPM characteristics at cryogenic temperatures}
\label{sec:charac}

\subsection{Pulse shape }
\label{sec:pulse shape}
\begin{figure}[htb!]
\centering 
\includegraphics[width=.65\textwidth,trim=0 5 0 40,clip]{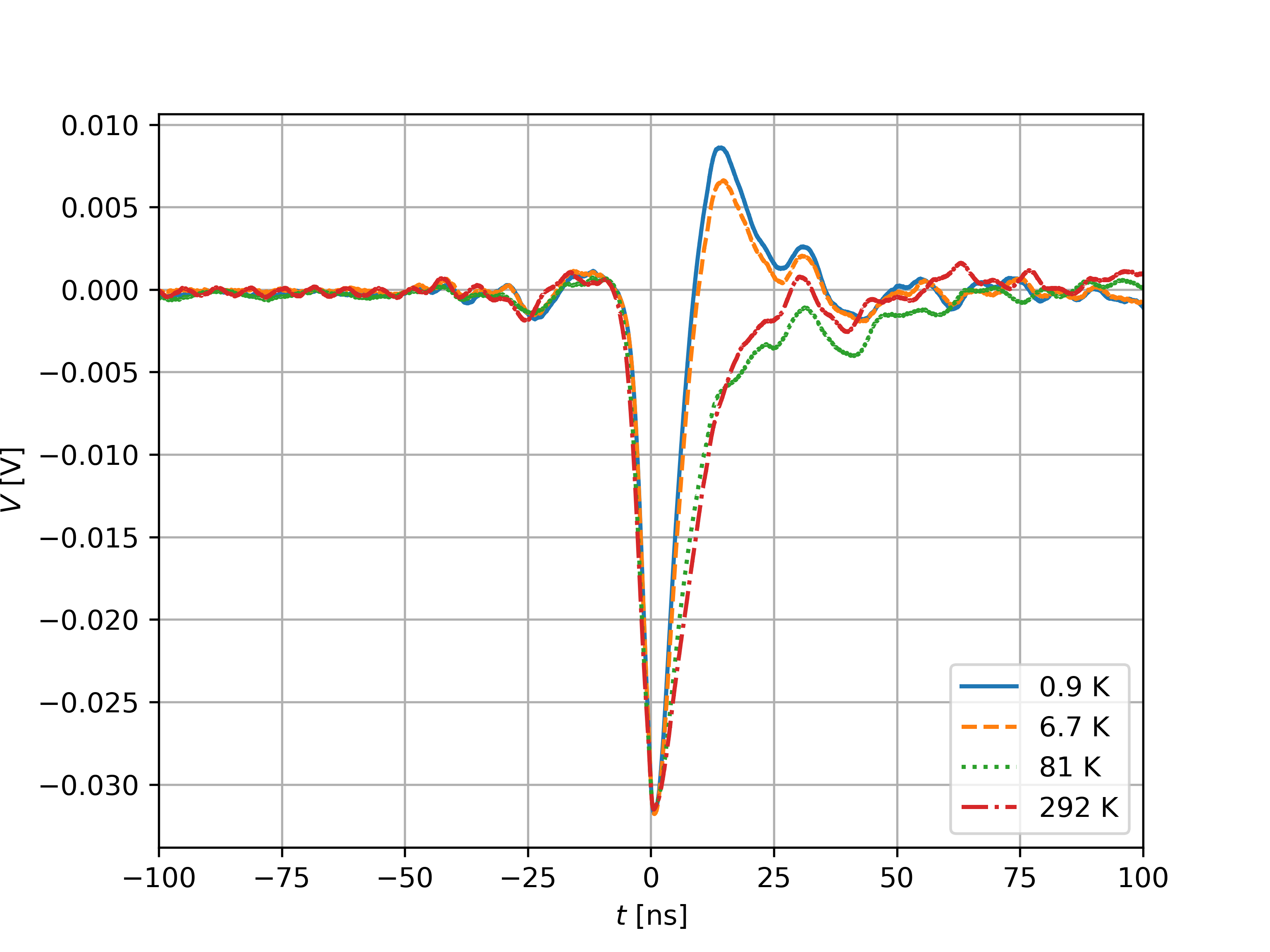}
\caption{\label{fig:waveform}
Averaged waveform at various temperatures with a constant overvoltage $\Delta V$ = 1.75 V. The fast rising edge is temperature-independent, while the slower falling edge becomes faster with increasing temperature. However, the overall pulse shape is stable over the whole temperature range. The overshoot is due to an impedance mismatch of the used micro coax cable at low temperatures.}
\end{figure}

In SiPMs, the pulse shape changes with temperature due to the temperature-dependent quenching resistance of the cell, which determines the pulse recovery time. Multiple signals under low-light conditions at various temperatures were recorded with the oscilloscope. In figure \ref{fig:waveform}, the average over 1000 waveforms is shown. It is visible that the fast rising edge of the pulse is not affected by the temperature change, where as the slower falling edge becomes faster at lower temperatures because the resistance of the metallic quenching resistor decreases with decreasing temperature. In the range from \SI{80}{\K} to \SI{300}{\K} the metallic quenching resistor of this SiPM has a good stability against temperature variations resulting in a stable pulse shape. Below \SI{7}{\K}, the pulse recovery time decreases by a factor of two but the overall pulse shape is stable.

\subsection{IV characteristics and breakdown voltage}
\begin{figure}[h!]
\centering 
\includegraphics[width=.48\textwidth,trim=10 0 10 0,clip]{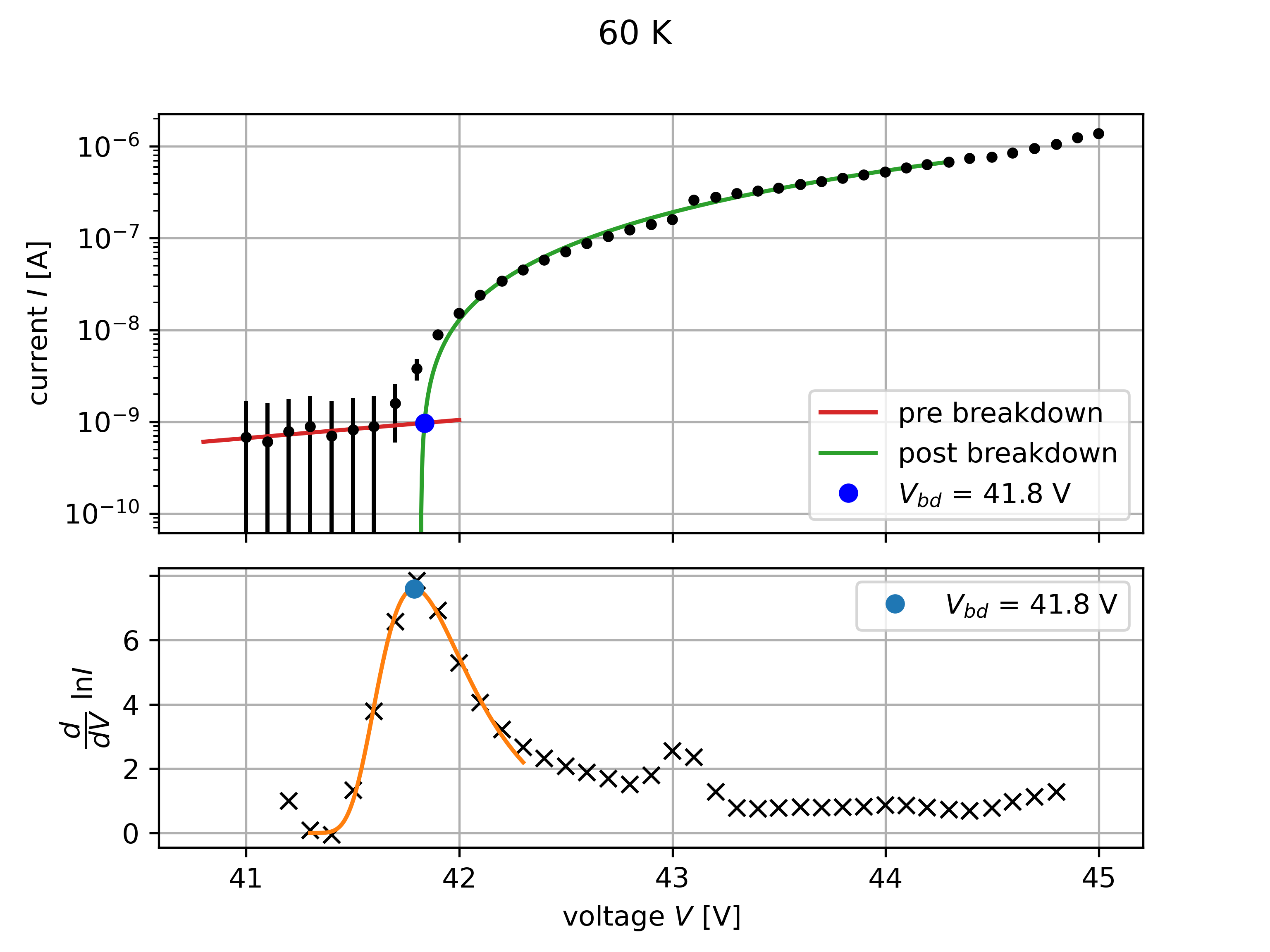}
\quad
\includegraphics[width=.48\textwidth,trim=10 10 10 10,clip]{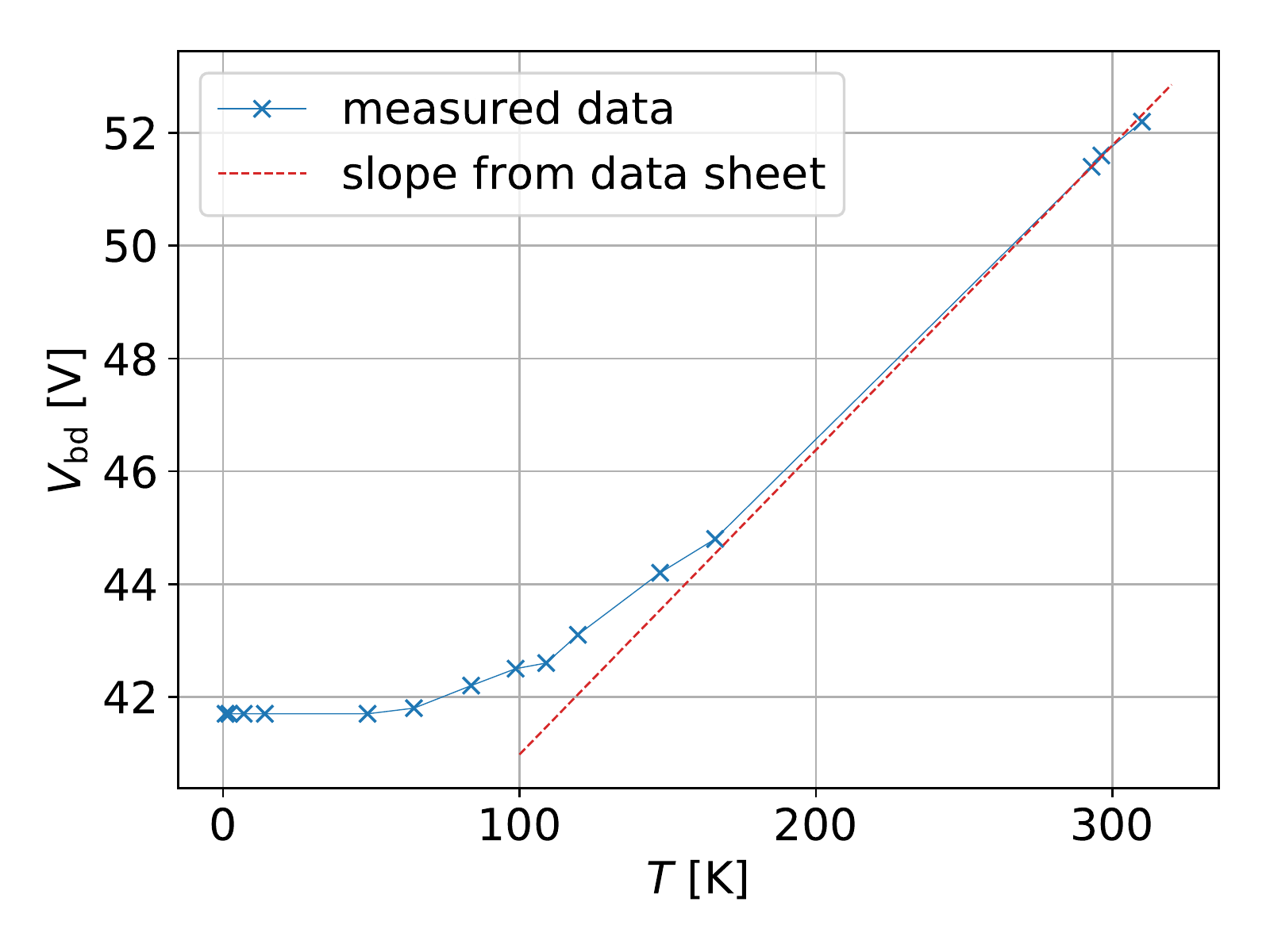}
\caption{\label{fig:breakdown}
Left: Breakdown voltage determination from the IV curve. The top shows the IV curve measured at \SI{60}{\K}, where two distinct regions corresponding to the pre (red) and post (green) breakdown are visible. A Landau peak is fitted on $\frac{\mathrm{d}}{\mathrm{d}V} \mathrm{ln}I$ to obtain the inflection point of the IV curve, which corresponds to the breakdown voltage. The discontinuity around 43 V is caused by the change of range of the measurement device. It lies outside of the fit region and did not influence the obtained result. Right: Breakdown voltage as a function of temperature. At lower temperatures a deviation from the linear dependence given by the datasheet is visible. }
\end{figure}

The reverse bias current-voltage (IV) curves of the SiPM were measured at temperatures ranging from \SI{0.85}{\K} to \SI{310}{\K}. At very low temperatures, both the leakage and avalanche current of the SiPM was in the range of the minimum instrument sensitivity. For this reason, the measurement of the IV curve was conducted under low-light condition, which allowed a reliable extraction of the breakdown voltage $V_\mathrm{bd}$. In figure \ref{fig:breakdown}, left, a typical IV curve recorded at $T = $ \SI{60}{\K} is shown. The shape of the curve is similar to one at room temperature: up to $V_\mathrm{bd}$ the reverse current increases gradually due to the dark current followed by a sudden increase at $V_\mathrm{bd}$ \cite{Chmill2017}. The breakdown voltage is extracted from the reverse IV curve as the voltage where $\frac{\mathrm{d}}{\mathrm{d}V} \mathrm{ln}I$ is maximal, corresponding to the inflection point of the IV curve as described in \cite{Otte2017}. The maximum of $\frac{\mathrm{d}}{\mathrm{d}V} \mathrm{ln}I$ was obtained by fitting a Landau distribution, which is used due to the skewed shape of the curve. 

The breakdown voltage measured at various temperatures is shown in figure \ref{fig:breakdown}, right. Between \SI{150}{\K} and \SI{300}{\K}, a linear decrease with temperature is visible, as reported in \cite{Baudis2018, Iwai2019}. The dashed red line matches the linear temperature coefficient of 54 mV/K around room temperature specified by Hamamatsu \cite{Hamamatsu}. A departure from the linear dependence is visible at lower temperatures, which can be explained by the temperature-dependent avalanche formation of charge carriers in silicon \cite{Crowell1966}.

\subsection{Operating range}
\begin{figure}[htb!]
\centering 
\includegraphics[width=.68\textwidth,trim=10 10 10 10,clip]{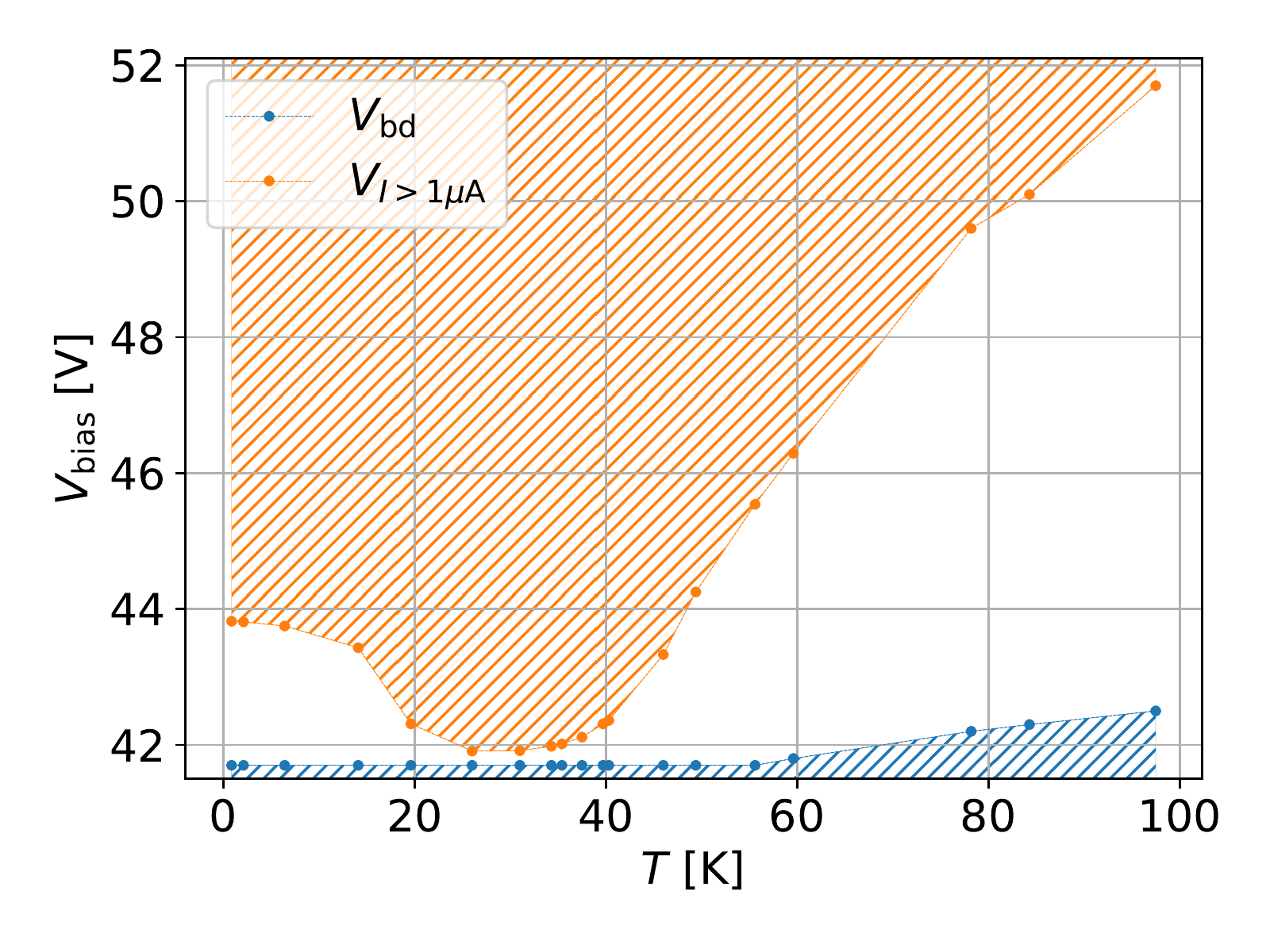}
\caption{\label{fig:operating range}
The useful SiPM operating range for single p.e. detection. The unshaded area represents the region where the SiPM can be operated in Geiger-mode within our dilution cryostat. The lower limit of $V_\mathrm{bias}$ is given by $V_\mathrm{bd}$ (blue), while an upper limit (orange) is defined by the voltage $V_{I>\SI{1}{\uA}}$ where the maximum tolerable dark current $I_{\max}>\SI{1}{\uA}$ was measured. Between \SI{20}{\K} and \SI{40}{\K} the SiPM cannot be operated reliably, while at lower temperatures operation with limited $V_\mathrm{bias}$ becomes possible again.}
\end{figure}

The operating range of the SiPM for single photoelectron (p.e.) detection was studied at various temperatures by monitoring the reverse current as a function of the bias voltage $V_\mathrm{bias}$. At cryogenic temperatures, dark current is limited, but delayed afterpulses are present due to the temperature-dependent trapping and release of charge carriers \cite{Ridley1983}. With higher $V_\mathrm{bias}$ the probability for afterpulsing increases \cite{Gallina2019}, which can mask the single photon signal. Further, this increases the current drawn by the SiPM resulting in heat dissipation. We defined an upper limit of a useful operational range by a maximum dark current $I_{\rm max} = \SI{1}{\uA}$ and the corresponding bias voltage $V_{I>\SI{1}{\uA}}$, at which the heat dissipation was tolerable in our dilution cryostat. We defined the lower limit of the SiPM operating range by the breakdown voltage. Figure \ref{fig:operating range} shows this above defined operating range for different temperatures between \SI{0.8}{\K} and \SI{100}{\K} as the unshaded region. This operating range becomes narrow when the temperature approaches \SI{40}{\K} due the increasing afterpulsing probability. In the temperature range between \SI{20}{\K} and \SI{40}{\K}, the increased afterpulsing probability caused self-sustaining afterpulsing processes, which narrowed the operational range to below 0.5 V. Below \SI{15}{\K} however, $V_{I>\SI{1}{\uA}}$ was observed to increase again. An explanation to this may be the increasing trapping time of the charge carriers with the decreasing temperature, a similar phenomenon already reported \cite{Cardini2014} at slightly higher (\SI{5}{\K}) temperatures. With the resulting large delays in afterpulsing, the useful operational range thus allowed to increase again, and the operation of the SiPM becomes possible with limited a overvoltage $\Delta V  = V_\mathrm{bias} - V_\mathrm{bd}\sim$ \SI{2}{\V} for {$T<15$}~K.

%\cite{Hornbeck1955}

\subsection{Photon detection efficiency}
\begin{figure}[h!]
\centering
\includegraphics[width=.5\textwidth,trim=0 0 0 0,clip]{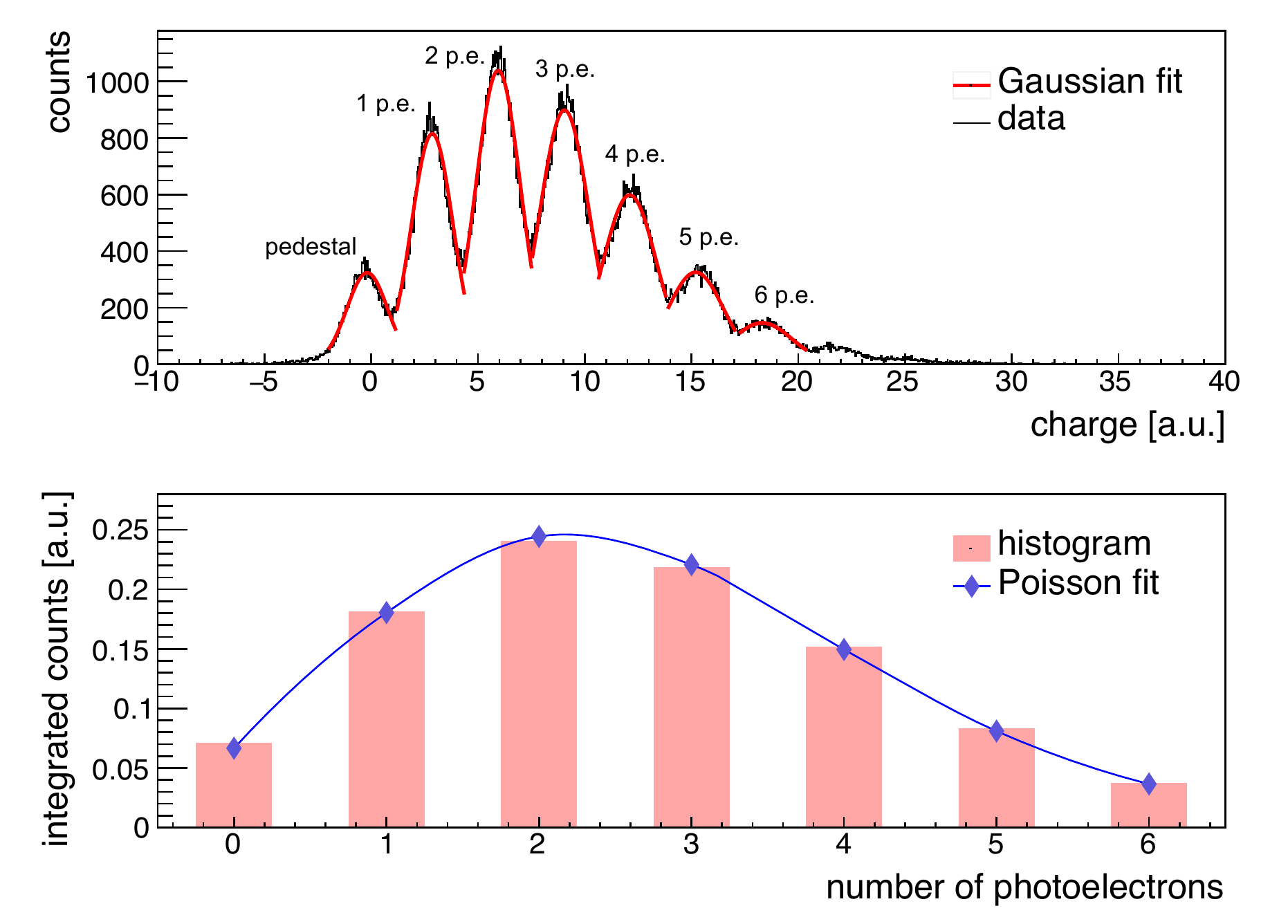}
\quad
\includegraphics[width=.42\textwidth,trim=15 12 15 18,clip]{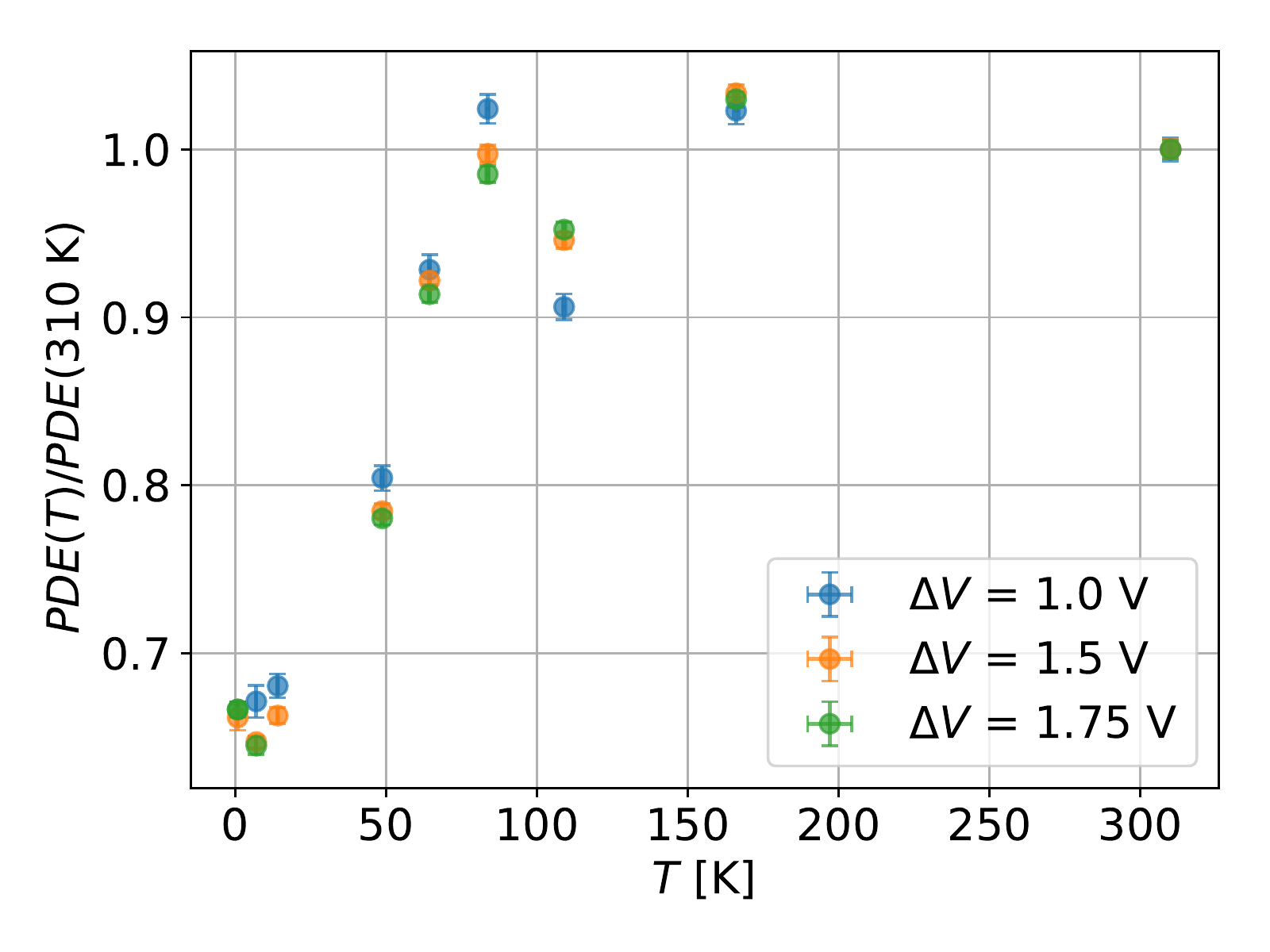}
\caption{\label{fig:PDE vs T}
Left: The top shows the SiPM charge spectrum measured at \SI{0.85}{\K} and $\Delta V = \SI{1.75}{\V}$, and the Gaussian fits on each p.e. peak. The bottom plot shows the entries under each peak, which is fitted by a Poisson function to obtain an average of 2.70 $\pm$ 0.02 p.e..  Right: Relative PDE compared to room temperature for various overvoltages. At \SI{0.85}{\K} the PDE is 0.666 $\pm$ 0.005 of the PDE at room temperature.}
\end{figure}

Single photon detection efficiency (PDE) at cryogenic temperatures at various $\Delta V$ was studied by illuminating the SiPM with light from the pulsed LED, resulting in a few photons reaching the sensitive area simultaneously. For each light pulse, the corresponding signal waveform was acquired and evaluated using the WaveDREAM board. The baseline was calculated from the first \SI{60}{\nano \s} of the waveform, and then subtracted from the subsequent raw waveform data. The charge spectrum was obtained by calculating the integral of the waveform within a time window of \SI{20}{\nano \s}. The first peak of the charge spectrum corresponds to the pedestal, where no photons are detected, and the other peaks from 1 p.e., 2 p.e., etc. signals. The peaks were fit with a Gaussian function to obtain the center of each peak. The distribution of the integrated entries under each peak was then fit by a Poisson function to obtain the average number of detected p.e.. Figure \ref{fig:PDE vs T}, left, shows the charge spectrum obtained a \SI{0.85}{\K} with $\Delta V$ = 1.75 V and the Poisson fit used to obtain the average number of detected photoelectrons.

The absolute PDE could not be determined since the absolute number of photons reaching the SiPM per LED pulse is not known. The relative PDE compared to room temperature was obtained by comparing the average number of detected p.e. from the same source at cryogenic temperature to the one at room temperature. Figure \ref{fig:PDE vs T}, right, shows the relative PDE at various temperatures for different $\Delta V$. Variations of the PDE are visible at temperatures above \SI{80}{\K}, while below \SI{80}{\K} the PDE is dropping. The PDE is given by the product of three factors: the geometric fill factor (ratio of the sensitive area to the SiPM area), the photo-conversion efficiency, and the avalanche triggering probability. As in \cite{Collazuol2011}, two counteracting mechanisms are expected to change the PDE with decreasing temperature. On one hand, the avalanche triggering probability is increasing with decreasing temperature due to the enhanced mobility of charge carriers, on the other hand, the energy gap is wider at lower temperatures, which decreases the photo-conversion efficiency. These two effects could explain the PDE variations above \SI{80}{\K}. The enhancement of the PDE at low $T$ is also observed \cite{Rech2007} for single avalanche photo diodes (APDs). Furthermore, at low temperatures charge carriers freeze out, resulting in carrier losses, which could explain the drop of PDE below \SI{80}{\K}. The PDE variation with temperature is also happening at other wavelengths as reported in \cite{Anfimov2021}.
At very low temperatures below \SI{1}{\K}, the PDE is a factor of 0.666 $\pm$ 0.005 smaller compared to room temperature, where the PDE is specified by the manufacturer to be at 40 \% for \SI{500}{\nano \m} light and $\Delta V$ = \SI{4}{\V} \cite{Hamamatsu}. Although, the SiPM at ultra low temperatures cannot be operated at such high gain conditions, single photon detection is nevertheless feasible at temperatures below \SI{1}{\K}.

\section{Commissioning of cryogenic scintillation detectors}
\label{sec:commissioning}
In the previous section we have shown that the Hamamatsu S13370 SiPM can be operated at temperatures below \SI{1}{\K}. To observe the formation of a M beam from a SFHe target, charged particle detection with finely segmented detectors operating at these ultra low temperatures is required. We developed cryogenic scintillation detectors based on the SiPMs characterized in Section \ref{sec:charac}. The SiPMs were directly coupled to plastic scintillator bars (Eljen EJ-204) which were wrapped with PTFE to improve light collection and reduce optical crosstalk between the detector channels. Since this type of SiPM is not covered with a protective layer, special care was required not to damage the wire bonds and silicon substrate. Coupling using optical cement damaged multiple detectors during the cooling-warming cycles such that we decided to use 3D-printed acrylic sleeves to fix the scintillators to the SiPMs.   

In total, we constructed 16 detectors consisting of plastic scintillator bars of the sizes 1.5 $\times$ 3 $\times$ 20 \si{\mm^3},  2 $\times$ 3 $\times$ 20 \si{\mm^3}, and  4 $\times$ 3 $\times$ 20 \si{\mm^3}, coupled to SiPMs with 3 $\times$ 3 \si{\mm^2} active area. The 16 detectors were placed on the first heat shield of the cryostat (position B in figure \ref{fig:Setup}) at a temperature of \SI{1.7}{\K}. A scintillator pill of the size of 5 $\times$ 6 $\times$ 6 \si{\mm^3} was coupled to a SiPM with 6 $\times$ 6 \si{\mm^2} active area and was placed within the target chamber at \SI{170}{\milli \K} (position A in figure \ref{fig:Setup}).

\begin{figure}[htb!]
\centering
\includegraphics[width=.4\textwidth,trim=2 2 2 0,clip]{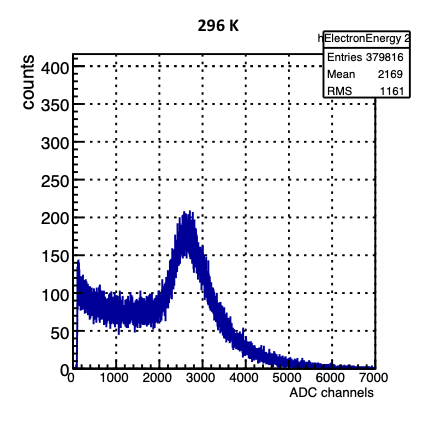}
\quad
\includegraphics[width=.4\textwidth,trim=2 2 2 0,clip]{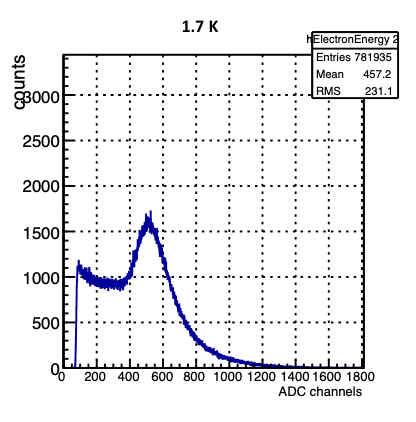}
\caption{\label{fig:energyspec}
Typical energy spectrum of positrons passing a thin plastic scintillator at two operation temperatures (Left):  \SI{296}{\K}, (Right): \SI{1.7}{\K}. In both cases, the full Landau peak, corresponding to the energy deposition of charged particles in a thin layer, is visible.}
\end{figure}

The detectors were commissioned using the muon beam from the $\pi$E1 area of the Paul Scherrer Institute. All detectors were capable of detecting positrons resulting from muon decays at the given temperatures. We did not observe a degradation in gain or change of breakdown voltage with multiple heat cycles between room temperature and the base temperature. The detector operated reliably and reproducibly over a data taking period of three weeks. Typical energy spectra of one of the 16 detectors at room temperature and at \SI{1.7}{\K} are shown in figure \ref{fig:energyspec}. The detected positrons have an initial energy distributed below \SI{50}{\MeV} and deposit part of their energy in a thin layer of scintillator material described by a Landau distribution. Based on Monte Carlo simulations, the Landau peak is situated at \SI{0.5}{\MeV} for muon-decay positrons traversing scintillator material with geometries described above. At the lower temperature, the measured Landau peak is situated at a lower ADC count number, indicating a lower gain of the detector. Nevertheless, the full peak is visible, showing the charged particle detection feasibility of the detector at ultra low temperatures.

\section{Conclusion}
\label{sec:Conclusion}
We have shown that SiPMs of the S13370 series from Hamamatsu can be operated in a dilution cryostat reaching temperatures below \SI{200}{\milli \K}. We measured the temperature dependence of various SiPM characteristics down to \SI{0.85}{\K}. The pulse shape has been shown to be stable with a decrease of the pulse recovery time at temperatures below \SI{7}{\K}. The breakdown voltage was determined using the reverse IV curves and shows a departure from the linear dependence with temperature given by the data sheet. The operating range was determined, where we have seen that in the temperature range from \SI{20}{\K} to \SI{40}{\K} the SiPM cannot be operated because afterpulsing effects cause the SiPM to draw too much current. We measured the temperature dependence of the relative PDE and determined the PDE at \SI{0.85}{\K} to be a factor of 0.666 $\pm$ 0.005 smaller than the one at room temperature, demonstrating the single photon detection capability at temperatures below \SI{1}{\K} with the studied SiPM. We developed finely segmented scintillation detectors using plastic scintillators and SiPMs, which we commissioned with a muon beam. In total, 16 channels of detectors at \SI{1.7}{\K} and one channel at \SI{170}{\milli\K} were operated reliably within a dilution cryostat and were capable of detecting charged particles.

%\appendix
%\section{Some title}
%Please always give a title also for appendices.

\acknowledgments
The authors would like to thank the HIPA facility at PSI for the stable beam and technical support. Special thanks to the PSI detector group, M.~Hildebrandt, A.~Stoykov, and F.~Barchetti and U.~Greuter and U.~Hartmann from the electronics group. This work was supported by the SNF Ambizione Grant No. PZ00P2 185975.

%\paragraph{Note added.} This is also a good position for notes added after the paper has been written.

\bibliographystyle{JHEP}
\bibliography{refs}

\end{document}